\documentclass[%
aip,
rsi
 amsmath,amssymb,
 reprint,%
]{revtex4-1}

\usepackage{graphicx}
\usepackage{dcolumn}
\usepackage{bm}
\usepackage{xcolor}
\usepackage[utf8]{inputenc}
\usepackage[T1]{fontenc}
\usepackage{mathptmx}
\usepackage[switch, modulo]{lineno}

\usepackage{tabularx,booktabs}
\newcolumntype{Y}{>{\centering\arraybackslash}X} 
\usepackage{array}

\begin{document}

\preprint{AIP/123-QED}

\title{A tunable dielectric resonator for axion searches at 11\,GHz}
\author{R.~Di Vora} \email{divora@pd.infn.it} 
\affiliation{INFN, Laboratori Nazionali di Legnaro, Legnaro, Padova, Italy}
\author{A.~Gardikiotis} \email{antonios.gardikiotis@cern.ch}
\affiliation{INFN, Sezione di Padova, Padova, Italy}
\author{C.~Braggio}\email{caterina.braggio@unipd.it}
\affiliation{INFN, Sezione di Padova, Padova, Italy} 
\affiliation{Dipartimento di Fisica e Astronomia, Padova, Italy}
\author{G.~Carugno} \affiliation{INFN, Sezione di Padova, Padova, Italy}
\author{A.~Lombardi} \affiliation{INFN, Laboratori Nazionali di Legnaro, Legnaro, Padova, Italy}
\author{A.~Ortolan} \affiliation{INFN, Laboratori Nazionali di Legnaro, Legnaro, Padova, Italy}
\author{G.~Ruoso} \affiliation{INFN, Laboratori Nazionali di Legnaro, Legnaro, Padova, Italy}

\date{\today}

\begin{abstract}
In the context of axion search with haloscopes, tunable cavity resonators with high quality factor and high effective volume at frequencies above about 8\,GHz are central for probing the axion-photon coupling with the required sensitivity to reach the QCD axion models. 
Higher order modes in dielectrically-loaded cavities allow for higher effective volumes and larger quality factors compared to basic cylindrical cavities, but a proper cavity frequency tuning mechanism to probe broad axion mass ranges is yet not available. \\
In this work, we report about the design and construction of a tunable prototype of a single-shell dielectric resonator with its axion-sensitive pseudo-TM$_{030}$ high-order mode at about 11\,GHz frequency.
A clamshell tuning method previously tested for empty cylindrical resonators has been perfected for this geometry through simulations and in situ tests conducted at cryogenic temperature. Tuning is accomplished in a range of about 2\,$\%$ the central frequency, without significantly impacting the quality factor of about 175000.
The experimental results presented in this work demonstrate the tunability of this type of resonator, definitely confirming its applicability to high frequency axion searches. 

\end{abstract}

\maketitle

\section{Introduction}

In the late 1970s, a new theory in particle physics was proposed to solve the strong CP problem  \cite{Peccei:1977,Peccei:1977ur}, a still open fundamental question about the violation of CP-symmetry expected in quantum chromodynamics (QCD), but never observed in any experiment involving only the strong interaction. 
The theory introduces a new symmetry and a scalar field which spontaneously breaks the symmetry, giving rise to a new fundamental particle named the axion \cite{Weinberg:1978,Wilczek:1978}. Researchers later realised that if these particles had low masses, in the range $1\,\mu$eV to 10\,meV, they could also be dark matter constituents \cite{PRESKILL1983127,Abbott:1983,IRASTORZA201889,Chadha-Day:2022}. 
In this mass range, the existence of axions as exclusive constituents of the galactic DM halo is probed by the haloscope detector, proposed by Sikivie in 1983 \cite{Sikivie:1983ip}.
A haloscope employs an electromagnetic resonant cavity immersed in a strong DC magnetic field coupled to a low-noise readout chain. The magnetic field acts as a source of virtual photons that couple with axions and convert them into real photons with frequency: 

\begin{equation}\label{eq:hf}
hf_{a}=m_{a}c^2+1/2 m_{a}v_{a}^2
\end{equation}
 
set by an axion mass $m_{a}$, c is the speed of light, $v_{a}$ the axion velocity in our galaxy and $h$ is the Planck’s constant. The expected velocity dispersion of axions is on the order of $10^{-3}\,c$, resulting in an effective axion quality factor $Q_{a} \sim 10^{6}$ in the laboratory frame \cite{Turner:1986aa}. 
A number of experiments worldwide use the cavity-based approach to search for these hypothetical particles by their interaction with the photon in the broadest possible mass range, and the related exclusion plot is updated by each experiment in specific frequency ranges \cite{cajohare}.  

If the cavity frequency $f_c$ matches the axion frequency $f_{a} \approx  m_{a}c^2/h$, in the limit $Q_0<Q_a$, the signal power in the cavity is given by:

\begin{equation}\label{eq:Power_axion}
P_{a\gamma\gamma}=g^2_{a \gamma\gamma}\frac{\rho_{a}}{m_{a}}{\omega}_c B_0^2VC_{mnl}Q_0\frac{\beta}{(\beta+1)^2}
\end{equation}

where $g_{a\gamma\gamma}$ is the axion-to-photon coupling, $\rho_a\cong 0.45\,{\rm GeV/cm^3}$ is the local halo DM density, $m_a$ is the axion mass, $B_0$ is the external magnetic field amplitude, $V$ and $Q_0$ are respectively the cavity volume and the unloaded quality factor of the axion-sensitive mode. $\beta$ is the coupling coefficient to the cavity mode of the receiver antenna. 
$C_{mnl}$ is the normalized form factor related to the spatial overlap between the external magnetic field and a given cavity mode labelled by the integer numbers $m,\,n,\,l$:

\begin{equation}\label{formfactor}
C_{mnl} = \frac{{|\int\vec{B}}\cdot\vec{E_{mnl}}dV|^2}   {B_0^2V\int_V{\epsilon_r}|E|^2dV},
\end{equation}

where $E_{mnl}$ is the electric field of the mode under consideration, and $\epsilon_r$ is the relative dielectric constant of the media inside the cavity. 

Even in haloscopes equipped with state-of-the-art cavites and SC magnets, $P_{a\gamma\gamma}$ of QCD axions is very small, well below $10^{-23}\,$W at 10\,GHz frequency. As the signal is much smaller than the noise introduced by the ultra low noise receivers employed for readout of the cavity signal, 
the signal to noise ratio needs to be improved by integration according to the Dicke radiometer equation. 
The most important figure of merit for an haloscope is thus the scan rate $df/dt$  \cite{Stern:2015}, namely the maximum speed at which it can probe different masses at a given sensitivity, given by

\begin{equation}\label{eq:scan_rate}
\frac{df}{dt}\varpropto \frac{1}{\rm SNR^2}\frac{g_{a\gamma\gamma}^4\rho_{a}^2Q_{a}}{m_{a}^2}\frac{B_0^4}{(k_BT_S)^2}C_{mnl}^2V^2Q_L
\end{equation}

where SNR is the target signal-to-noise ratio of the experiment, $k_B$ is the Boltzmann constant, and $T_{S}$ is the total system noise temperature, which represents the random Nyquist noise in the system in terms of an equivalent effective temperature. $Q_L$ is the loaded quality factor of the cavity given by the relation $Q_L=Q_0/(1+\beta)$. 
The cavity-dependent contributions to the scan rate can be collected in the figure of merit $\mathcal{F}=C_{mnl}^2 V^2 Q_0$, which needs to be maximised for every haloscope experiment. 

Extending the search to higher axion masses with cavity-based haloscopes is challenging as the cavity radius needs to be approximately $\lambda_a/2$ where $\lambda_a=h/m_ac $ is the Compton wavelength given by the axion mass. As the resonant frequency increases, the cavity volume decreases accordingly, resulting in largely reduced sensitivities. 
For this reason, dielectric haloscopes as ORPHEUS \cite{Rybka2015,Cervantes2022}, MADMAX \cite{Caldwell2017,Knirck_2021} and the ALPHA plasma haloscope \cite{Lawson2019,Millar2023}, have been proposed to overcome the limitations of the cavity-based approach for the search of heavier axions at relevant sensitivity. 

In this work, the high frequency challenge is addressed by exploiting higher order modes in dielectric resonators \cite{Kim:2019asb,McAllister:2018}. Previous efforts on the concept from our collaboration focused on a high Q dielectric resonator based on two dielectric shells, which demonstrated $\sim10^7$ quality factor at 8\,T field but was limited in sensitivity by its small form factor $C_{030}=0.03$\cite{DiVora2022}. Moreover, while tuning of the cavity was achieved over a small range, the implementation of a wider bandwidth tuning method represents an experimental challenge. 

Here, we test a tunable implementation of the dielectric cavity concept, where one dielectric shell is 
used to reshape the TM$_{030}$ mode profile and get a higher $C_{030}$ compared to the two shells geometry. The expected quality factor is in this case smaller, but anyway it exceeds the one  of cylindrical copper cavities.
A thorough study including finite element method simulations, bead pull measurements and mode maps at cryogenic temperature are conducted around a prototype resonator that, even though it has not been fully optimized, exhibits promising figures of merit.

\section{cavity design}

\begin{figure}
\begin{center}
\includegraphics[width=3.5in]{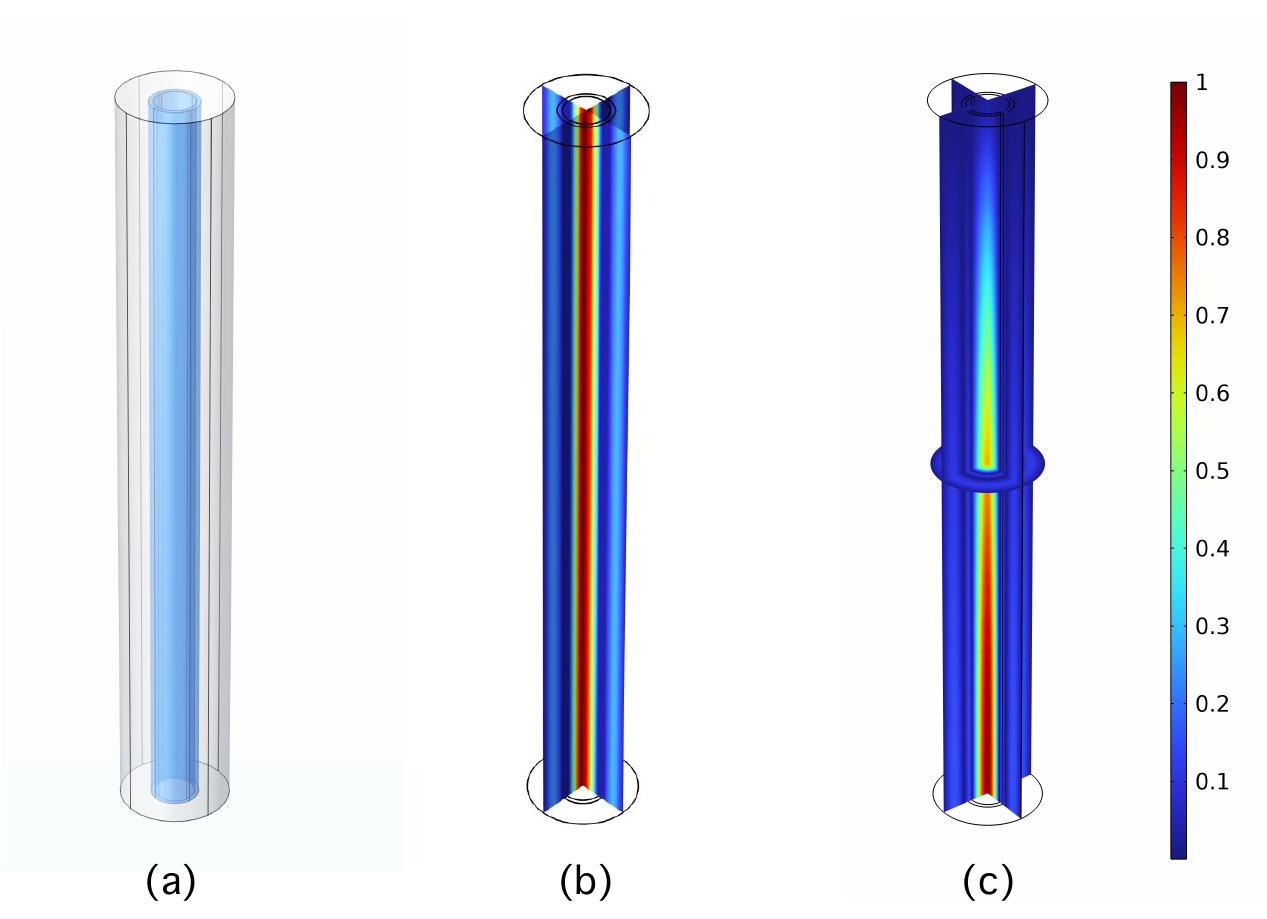}
\caption{(a) Single-shell dielectric cavity model with $L_c$= 422.2\,mm and radius $r_c$=30\,mm, and (b) electric field profile of its TM$_{030}$. (c) The longitudinal symmetry of the TM$_{030}$-mode is broken when a gap of about 2\,mm between the top endplate and the sapphire shell is considered.}
\label{fig:1}
\end{center}
\end{figure}

The conventional mode considered in cylindrical cavities in axion haloscopes is usually the TM$_{010}$ since it yields the highest form factor $(C_{010} = 0.69)$ of a given detection cavity volume in a solenoidal magnetic field. Higher order modes, while allowing higher cavity volume at equal frequency, have significantly smaller $C$ factors $(C_{020} = 0.13$ and $C_{030} = 0.05)$ due to the presence of out-of-phase field components.
In this paper we use a single dielectric hollow cylinder design (see Fig.\ref{fig:1}\,(a)) to suppress the negative radial lobe of the TM$_{030}$ mode. As an overall result, the effective volume $V_{eff}=C_{030}\,V$ of a dielectric cavity is larger than that of a TM$_{010}$ in the conventional cylindrical cavity. 
For a given frequency of resonance, in this design the figure of merit $\mathcal{F}$ is maximized by choosing the correct position and thickness of the dielectric shell. In particular, as reported in Ref.\,\cite{McAllister:2018}, in an empty cylindrical cavity the radial distance from the cavity axis to the point where the TM$_{030}$ mode electric field changes direction is $0.278\,r$, with $r$ cavity radius. The region of out of phase field is $0.362\,r$-thick, therefore the dielectric shell should have a thickness of $0.362\,r$\,/$\sqrt{\epsilon_r}$, with $\epsilon_r$ dielectric constant of sapphire. 

The realization of such a design for operation at cryogenic temperatures is rather challenging, especially if tuning is also to be accomplished. The first issue to address is related to the rather large difference between the thermal expansion coefficients of copper and sapphire. 
A gap is then needed at room temperature between the top of the sapphire shell and the top copper endcap, which then decreases in cryogenic conditions, with a residual empty space left to protect the sapphire shell. The effect of this gap is to significantly reduce the longitudinal symmetry of the mode of interest. As shown in Fig.\ref{fig:1}\,(c), the field mode is concentrated on the opposite side of the cavity when, for instance, a 420\,mm-long dielectric shell is concentric to a copper cylinder with radius $r_c=30$\,mm and 422.2\,mm-long. 

\begin{figure}[ht]
\includegraphics[width=0.48\textwidth]{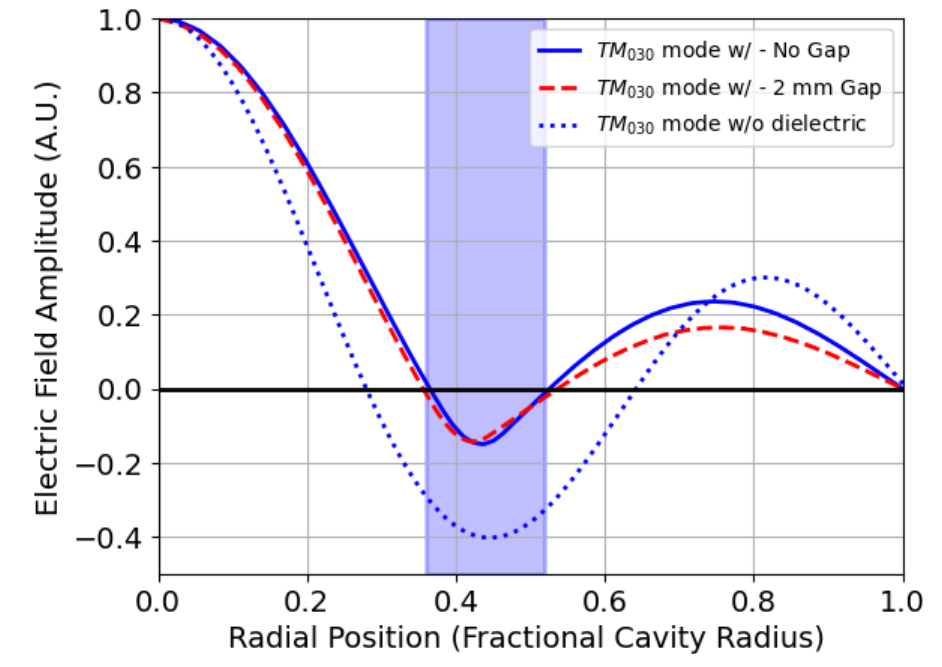}
\caption{\label{fig:2} Comparison between the electric field profile of the dielectric resonator TM$_{030}$ mode with (red dashed) and without a gap (blue line) between the top of the sapphire shell and the copper endcap. The gap influences the outer field lobe, giving rise to a smaller contribution to the $C$-factor (see eq.\,\ref{formfactor}). 
For comparison, we report also the analytical solution of the TM$_{030}$ mode inside an empty cavity (blue dotted line). Note that the maximum field strength and cavity radius are normalized to unity for each configuration, and that the fields are calculated at the longitudinal center of the cavity.}
\end{figure}

In addition, we observe that the in phase-component (see Fig.\,\ref{fig:2}) of the electric field in the region between the shell and the external copper cylinder has a smaller amplitude when the gap is present, and this in turn negatively impacts on $C_{030}$.
We thus want to minimize this gap in our realizations. 

The geometry that optimizes at a frequency of 11\,GHz the previously discussed out-of-phase lobe suppression using a dielectric shell in the TM$_{030}$ mode, has radius of the resonator $r=28.1\,$mm, and sapphire thickness $t=3.97\,$mm. 
Compared to an empty cylindrical cavity resonating at the same frequency, this dielectric resonator has an effective volume of $C_{030}V=0.48$ greater by a factor 5. 

\section{tuning mechanism} 
Typical tuning mechanisms in cylindrical cavities involve the movement of metallic rod(s) that break the transverse symmetry. Moving post patterns in low resonant frequency cavities might be effective, but at higher frequencies this procedure is either impractical or reduces the sensitivity \cite{Simanovskaia:2021}.

In this work, we rely on a previously introduced clamshell cavity tuning system\cite{Braggio:2023gnd}, whereby the copper cavity is divided into two halves, that can be radially opened while joined at a fixed edge to increase the effective radius and correspondingly decrease the frequency \cite{Kim:2019asb}. 
The tuning was tested in an empty copper cylindrical cavity at cryogenic temperatures, and a tuning range of 300\,MHz was demonstrated for the 10.3\,GHz-frequency TM$_{010}$ mode with quality factor $Q_0=5\times10^4$. 

Here, we employ for the dielectric resonator the same mechanism for opening the external copper cylinder, while the sapphire shell is kept in a fixed position. 
In the prototype cavity, the high purity, c-axis sapphire (Al$_2$O$_3$) hollow cylinder is held in place by Teflon pins, eliminating the need for grooves at the copper endplates. 
The Teflon pins used to keep the sapphire in place do not contribute to the overall cavity loss, and we avoid spurious modes arising from the grooves that would negatively impact the TM$_{030}$ mode quality factor.

In section \ref{fem_meas} we report a systematic study based on FEM simulations we did to understand the resonator performance while it is tuned, and compare the results with experimental measurements obtained with a prototype resonator with its axion-sensitive mode TM$_{030}$ at 11\,GHz. The investigated range is of about 70\,MHz, which can be reached when the opening angle of the two cavity shells is about one degree. 

RF gaskets are used to ensure a good RF contact to the static endplates while the cavity frequency is tuned, as they preserve the current pathways required to sustain the mode and in turn reduce radiation leak through the gaps. In dedicated simulations we verified that the gaskets do not significantly impact the cavity quality factor and can thus be modelled as bulk metal.

\begin{figure}
\begin{center}
\includegraphics[width=3in]{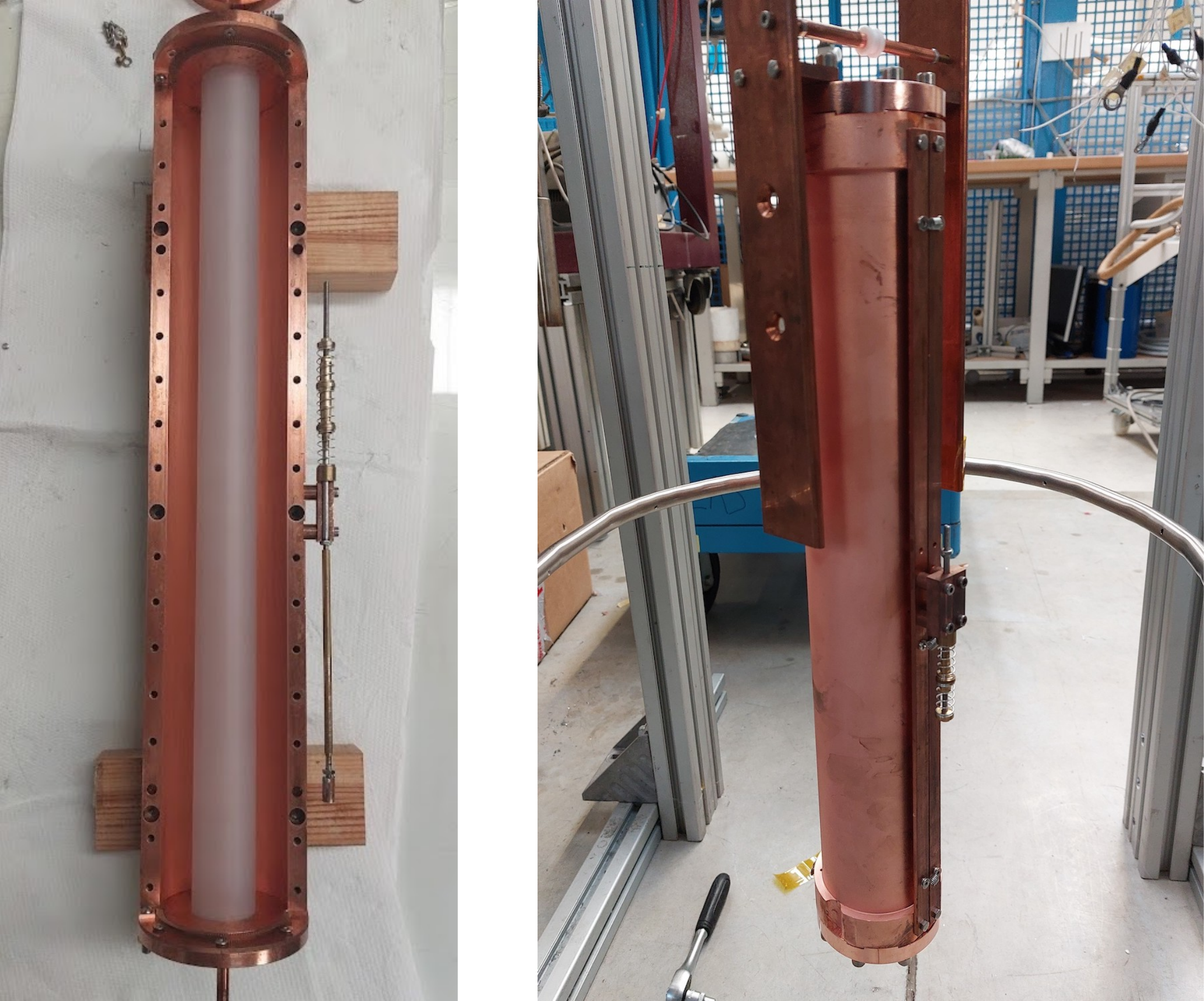}
\caption{(a) Open dielectric cavity exposure shows the sapphire hollow cylinder. (b) Cavity assembly for cryogenic measurements. The clamshell tuning mechanism attached on the right side is composed of two copper blocks and a conical wedge (not visible). The latter is inserted into the copper blocks and rotates the moving side of the cavity when pulled upward by a thin steel wire.}
\label{fig:3}
\end{center}
\end{figure}

The mechanism controlling the opening angle is automated using the controller of a high-resolution linear stepper motor at room temperature. The motor is connected to a conical wedge used to separate the two copper shells (see Fig.\,\ref{fig:3}) through a thin steel wire of approximately 2.2\,m length. The force applied by the wedge is balanced by four springs mounted next to it, and near the top and bottom cavity endcaps.

\section{Bead-pull measurements}\label{bead}

\begin{figure}[ht]%
\includegraphics[width=0.35\textwidth]{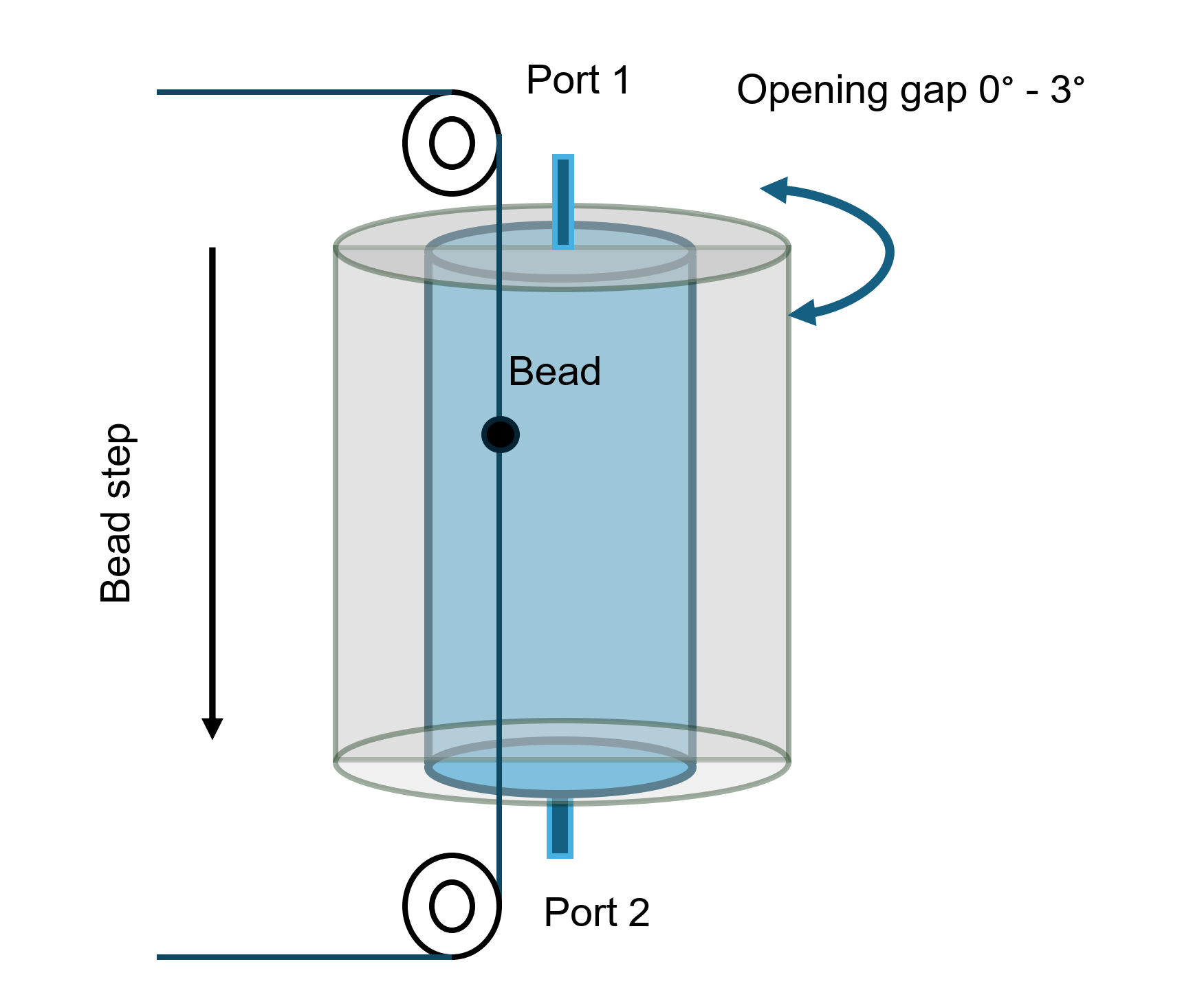}
\caption{\label{fig:6} Schematic representation of the bead-pull measurement conducted on the prototype cavity at room temperature. A pulley system allows to guide the Kevlar thread supporting the bead. We used this method to infer the electric field profile at different opening angles of the dielectric cavity.}
\end{figure}

The electric field intensity for the TM$_{030}$ mode, that is localized mainly inside the dielectric hollow, is constant along the cavity length in the ideal case.
However, tilts or misalignments of the cavity parts, as well as the mentioned gap issue, give rise to asymmetries in the axial direction and in turn to field concentration towards one end of the cavity. Additionally, the same nonidealities may in principle cause unwanted mode mixings as seen in Fig.\,\ref{fig:4}\,(b).
It is thus of central importance to map the electric field profile within the cavity to assess the real sensitivity of the haloscope. 
To measure the electric field profile distribution along the longitudinal direction, we use the “bead-pull” technique commonly employed in accelerator physics \cite{Maier1952}. A bead-pull RF measurement entails pulling a small dielectric or metallic bead through a cavity while taking transmission measurements of the scattering parameter $S_{21}$. The method is based on the Slater perturbation theory, which states that if a cavity geometry is perturbed slightly, we can calculate the first-order shift $\Delta f$ in the eigenfrequency $f$ induced by the bead:

\begin{equation}\label{eq:BEAD}
 \frac{\Delta f}{f}= \frac{-(\epsilon-1)}
{2}\frac{V_{b}}{V_{c}}\frac{E(r)^2}{\langle E(r)^2 \rangle_{c}}
\end{equation}

where $V_b$ ($V_c$) is the volume of the bead (cavity) and $\epsilon$ is the dielectric constant of the bead. The shift of the resonant frequency is proportional to the strength of the squared electric field at the bead location.
The measurement setup is schematized in Fig.\,\ref{fig:6}. The bead is a $8.2\,$mm-height, $1.5\,$mm-diameter cylinder of Low-Density Polyethylene (LDPE) with $\epsilon$ = 2.24 at 11.2\,GHz\cite{Strt2008DielectricPO}. The bead is attached to a 80\,$\mu$m-thick Kevlar string, which longitudinally traverses the cavity at fixed radial and azimuthal position.

\begin{figure}[ht]%
\includegraphics[width=0.5\textwidth]{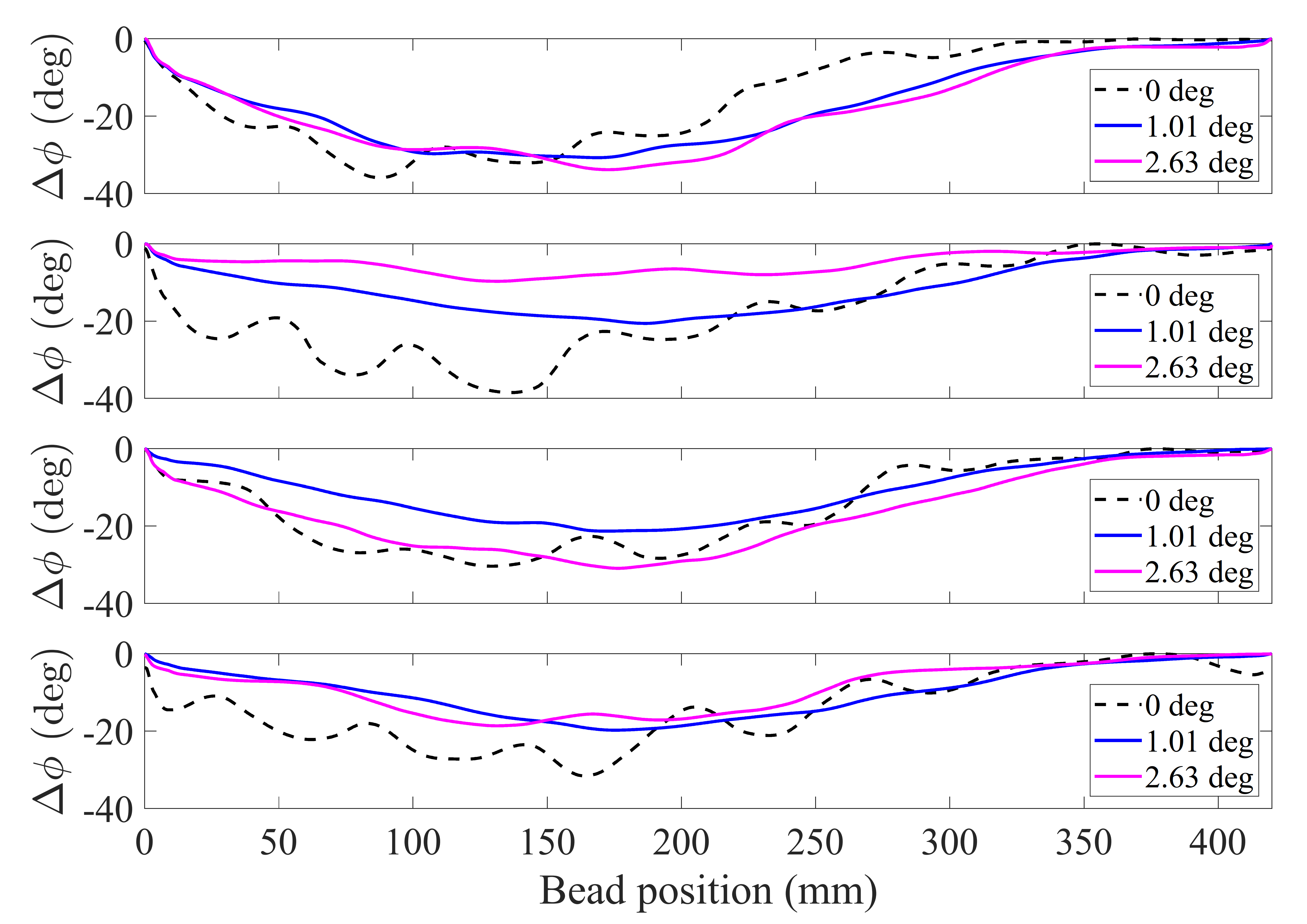}
\caption{\label{fig:7} Measured electric field phase shifts with the bead pull method. The shift is given at different positions along the cavity axis for three different clamshell openings. 
When the cavity is closed ($0^{\circ}$), the TM$_{030}$ mode is perturbed by an intruder TE mode, which determines the oscillating character of the field profile. We do not observe this oscillation when the cavity is open, as shown at 1.01$^{\circ}$ and 2.63$^{\circ}$ opening angles. In addition, we observe field mode concentration at the cavity bottom due to the gap at the top endplate. 
}
\end{figure}

We chose to record the phase shift $\phi(f_0)$ of the transmission parameter $S_{21}$ instead of the frequency shift as the former is easier to detect on a vector network analyzer. $\phi(f_0)$ is related to the frequency shift $\Delta f$ through the relation: 

 \begin{equation}\label{eq:BEAD1}
 \frac{\Delta f}{f}=\frac{\tan[\phi(f_0)]}{2Q_L},
\end{equation}

where $Q_L$ is the loaded quality factor of the unperturbed cavity. 
The results reported in Fig.\,\ref{fig:7} for three values of the opening angle (0°, 1.01° and 2.63°), with large frequency shifts recorded when the bead is close to the bottom of the cavity, clearly indicate mode localisation when a gap is present between the dielectric hollow and the endcap. 

Note also the oscillatory character of the hybrid mode present at 0$^{\circ}$ opening angle, as predicted by the simulations. 
The bead-pull measurements show that the axion-sensitive mode mixes with a mode with 7 longitudinal lobes, while at 1.01$^{\circ}$ and 2.63$^{\circ}$ opening angles the TM$_{030}$ pure mode character is observed.

\section{The dielectric cavity prototype: FEM simulations and experimental results} \label{fem_meas}

The prototype cavity was designed to resonate at about 11\,GHz, starting from a sapphire tube that was available in the laboratory. All the relevant dimensions are reported in Tab.\,\ref{table:1}.
\begin{table*}
    \begin{center}
    \begin{tabularx}{1\textwidth}
    {|>{\centering\arraybackslash}X 
    |>{\centering\arraybackslash}X 
    |>{\centering\arraybackslash}X 
    |>{\centering\arraybackslash}X 
    |>{\centering\arraybackslash}X 
    |>{\centering\arraybackslash}X 
    |>{\centering\arraybackslash}X 
    |>{\centering\arraybackslash}X 
    |>{\centering\arraybackslash}X |}

    \hline
         $T$\,(K)&  $L_{c}$ &  $r_{c}\,$ & $L_{Al_2O_3}\,$ & $r_{Al_2O_3}\,$ & $t$\, & $\epsilon$  &  $\tan\delta$ & $\sigma$\,(S/m)   \\
         \hline
        300 &  422.23  & 28.6   &  420&    10.875  &  2.35  & 11.44  &  $4.872\cdot 10^{-6}$  &  $5.8\cdot 10^7$     \\
         \hline
        4 & 420.39 & 28.475 & 419.66 & 10.866 & 2.348 &11.2 & $1.188\cdot 10^{-9}$ & $3.18\cdot 10^8$    \\
         \hline
    \end{tabularx}
\end{center}
    \caption{
    Dimensions measured at room temperature and calculated at 4\,K for the prototype cavity used in the simulations.
   $L_c$ is the cavity length, $r_{c}$ the internal radius of the copper shell, $L_{Al_2O_3}$ is the sapphire cylinder length with internal radius $r_{Al_2O_3}$ and thickness $t$. All the dimensions are given in mm. The values of the sapphire dielectric constant $\epsilon$, sapphire tangent loss $\tan\delta$, and copper conductivity $\sigma$ used in the simulations are also reported.}
\label{table:1}
\end{table*}

As shown in Tab.\,\ref{table:2}, even though this geometry is not optimized according to the previously reported rules, 
its figure of merit $\mathcal{F}$ largely exceeds that of a conventional pill-box cavity resonating with its fundamental mode at the same frequency (Tab.\,\ref{table:2}).

\begin{table}
\begin{ruledtabular}
\begin{tabular}{lcccc}
model &\mbox{$Q$}&\mbox{$C$}&$\mathcal{F}$\\
\hline
\\ dielectric (ideal) TM$_{030}$& 182000 & 0.476 & 41300 \\
w/o gap, $t=3.97$\,mm & & &\\
\hline
\\ dielectric (prototype) TM$_{030}$ & 328000 & 0.235 & 20000\\
w 0.73\,mm gap, $t=2.35$\,mm & & &\\
\hline
\\ pill-box TM$_{010}$ & 76000 & 0.667 & 700\\
\end{tabular}
\end{ruledtabular}
\caption{\label{table:2}
Comparison between the ideal, in which the sapphire thickness is $t = 0.107\,r_c$, $r_c$ is the copper cylinder radius\cite{McAllister:2018}, and the present prototype dielectric cavity models, with axion-sensitive mode TM$_{030}$ at 11\,GHz. In the ideal case, a null gap has been assumed, whereas the prototype model has a gap of 0.73\,mm, as is the case at cryogenic temperature. 
In addition, we report the cavity parameters for a conventional cylindrical cavity with TM$_{010}$ at the same frequency. 
For the ideal and pill-box models we considered 420\,mm-length copper cylinder. The prototype has 420.39\,mm copper cylinder length. Assumed copper conductivity for all models is $3.18\times10^{8}$\,S/m.}
\end{table}

\begin{figure}
  \includegraphics[width=0.425\textwidth]{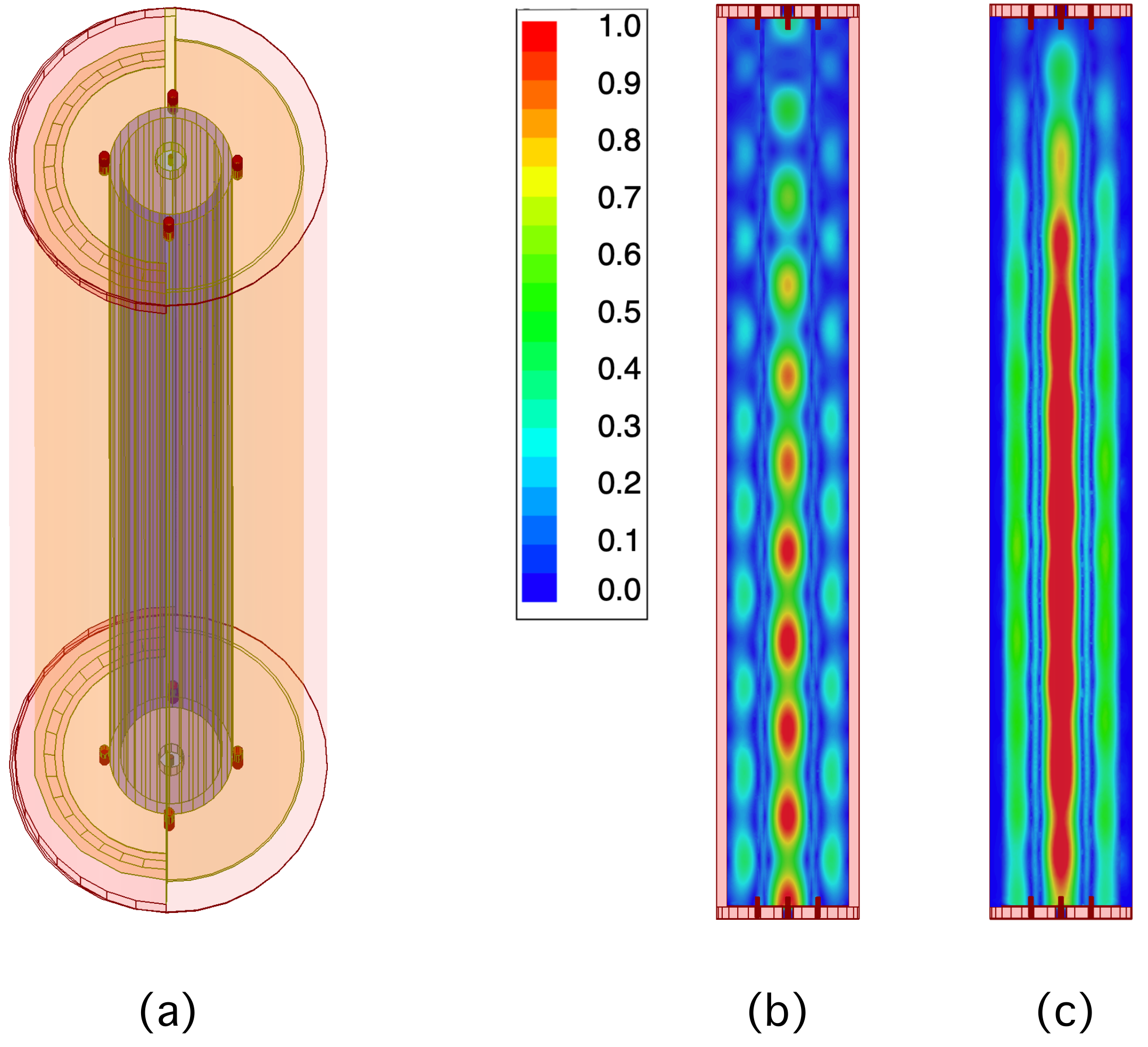}
  \caption{\label{fig:4} (a) Model of the prototype cavity used for the simulations with HFSS. Teflon pins (in red) hold the dielectric cylinder. Electric field profile of the TM$_{030}$ mode perturbed by an intruder mode at zero opening angle (a), and (b) effect of the gap at 1.2 degrees opening angle. Coupling antennas on both the top and bottom endplates are visible, at the cavity axis. }
\end{figure}

We perform FEM simulations based on a model having the measured dimensions of the prototype cavity (see Table \ref{table:1}), rescaled at liquid helium temperature by taking into account the materials thermal expansion coefficient \cite{simon1992nist,bradley2013properties}. 
The tangent loss of the sapphire shell is assumed, as well as its dielectric constant at cryogenic temperatures, respectively $1.18\cdot 10^{-9}$ and 11.2. Copper has been assigned a conductivity of 5.48 times its conductivity at room temperature ($\sigma=5.8\cdot 10^7$ S/m) as calculated considering the anomalous skin effect (ASE) \cite{Calatroni:2718002}. The scattering parameters are calculated using the driven modal solver to get reliable $C$-factors. As depicted in Fig.\,\ref{fig:4}\,(a), the strong (weak) port antenna on the top (bottom) endplate are included in the model. 

Fig.\,\ref{fig:4} shows that the field at a 0$^{\circ}$ opening is perturbed by a TM$_{mn0}$, $n=10$ intruder mode, while the profile at 1.2$^{\circ}$ looks much less perturbed. 

\begin{figure}[ht]%
\includegraphics[width=0.48\textwidth]{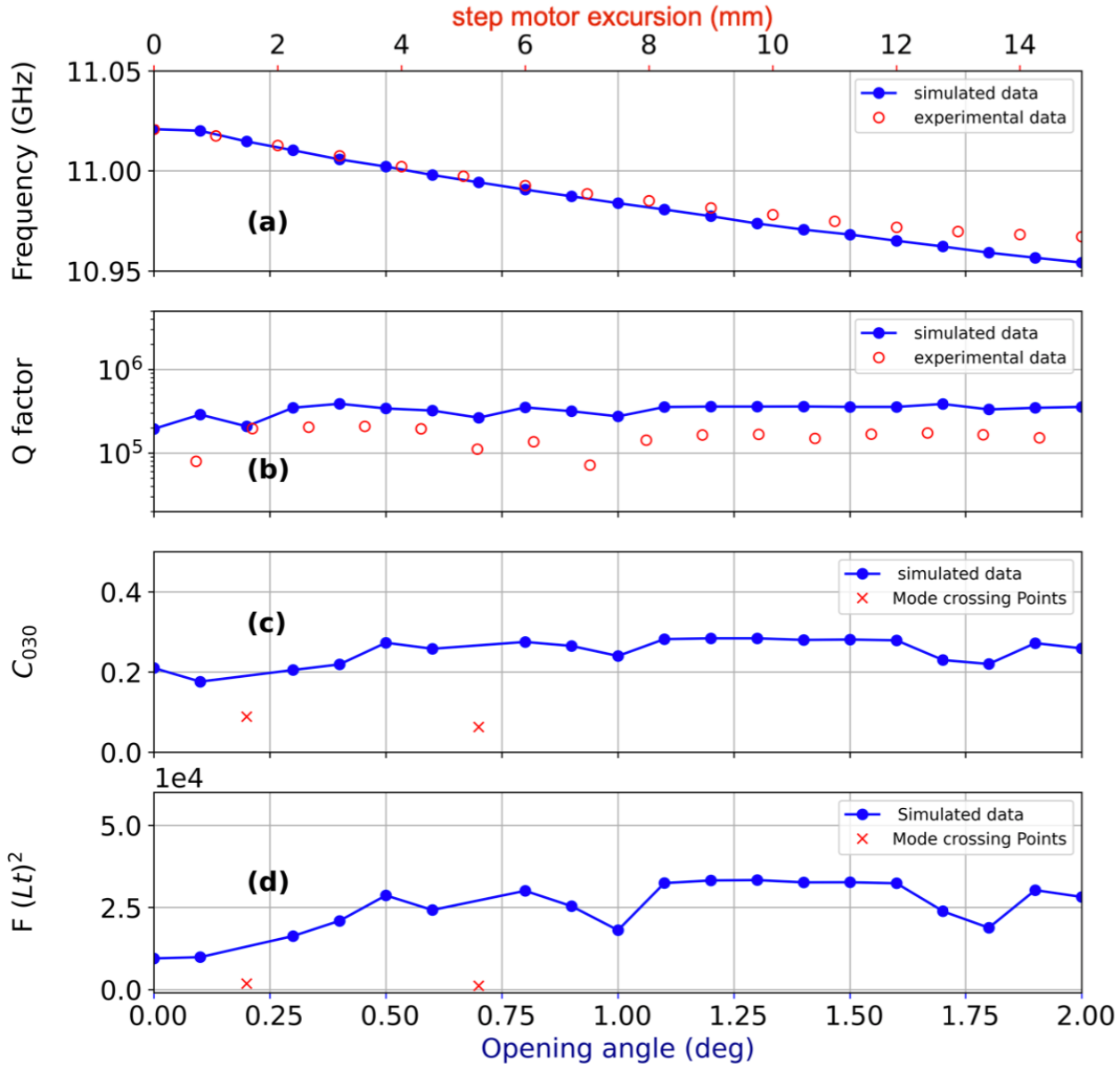}
\caption{\label{fig:5} (a) Frequency tuning of the dielectric cavity for different apertures: comparison between FEM simulation results and measurements obtained with the prototype cavity kept at at about 5\,K. (b) Same comparison for the unloaded quality factor. Plot (c) reports the calculated form factor $C_{030}$. The red cross points represent the frequency region with mode crossing, as well as in plot (d), where the figure of merit $\mathcal{F}=(CV)^2Q$ is given for different opening angles.}
\end{figure}

It is important to notice that at each opening angle the Q-factor can be inferred by fitting the transmission spectrum, even in case of perturbed modes. 
The simulations results are compared with the experimental data obtained with the prototype cavity cooled down to about 5\,K in Fig.\,\ref{fig:5}. 
Simulations were performed at steps of 0.1$^{\circ}$ between 0$^{\circ}$ and 2$^{\circ}$.

 By fitting the transmission and reflection spectra, we obtain the quality factor and the coupling coefficient $\beta$. As the relation between opening angle and tuning motor position is unknown, to compare the experimental data with the simulations results, we rescale the axes by imposing the same slope in the linear region to the corresponding data. In general, the behaviour of the dielectric cavity while it is tuned complies with what we expect by the simulations, both in terms of frequency and quality factor. Note that at large opening angles the measured frequency slightly deviates from the value obtained by the simulations due to the bottom-of-the-line behaviour of the tuning mechanism. A linear correspondence between steps and aperture angle is thus not ensured throughout the entire tuning range. 
 
The measured quality factor is constant for the probed opening angles, but lower than the simulation values by about a factor 2 probably due to the low conductivity at cryogenic temperatures of the phosphor bronze used for the RF gaskets. In addition, the pressure exerted by the gaskets at the contact points might not be sufficient to ensure a good RF contact.

As previously mentioned, when intruder modes approach the frequency of the axion-sensitive mode, the two modes mix, resulting in a hybrid mode (see Fig.\,\ref{fig:4}\,(b)). This in turn reduces the sensitivity in small frequency intervals within the overall tuning range, as can be seen from the red-cross data in Fig.\,\ref{fig:5}\, (c) and (d).

\begin{figure}
  \includegraphics[width=0.49\textwidth]{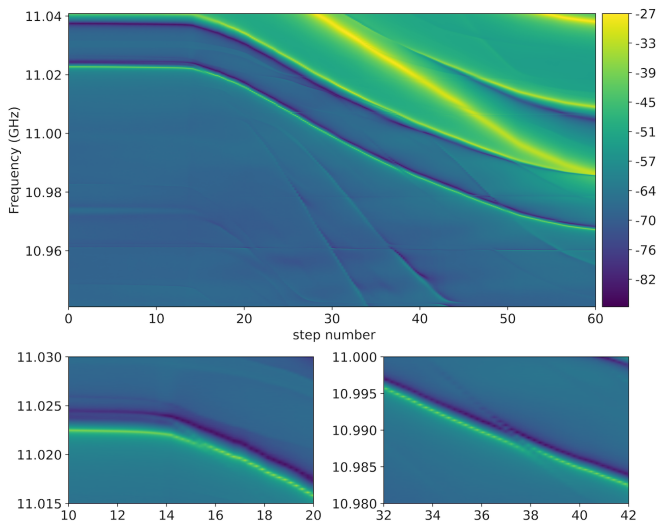}
  \caption{\label{Fig:8} Experimental mode map plot of the dielectric cavity prototype obtained at about 5\,K. Hysteresis of the tuning mechanism prevents variation of the aperture up to about 60 steps (arbitrary unit). (b) Mode crossing around the closed cavity position. (c) Avoided mode crossing. }
\end{figure}

The mode map reported in Fig.\,\ref{Fig:8} has been obtained at about 5\,K, opening the copper clamshell up to 2$^{\circ}$, corresponding to a frequency change of 60\,MHz for the TM$_{030}$ mode.
 A stepper motor is used to control and adjust the strong port antenna, ensuring critical coupling condition of the transmission line to the cavity mode at every frequency step.
The two adjacent modes TM$_{030}$ and the TM$_{031}$ change synchronously without significant mode mixing effects. 
Indeed, the TM$_{030}$ mode crosses few intruder modes indicated by the arrows in Fig.\,\ref{Fig:8}, but they reduce the sensitivity in a limited region spanning few \% of the full tuning range. Note also that the TM$_{030}$ mode frequency decreases more slowly compared to the intruder modes.

\section{Discussion and conclusions}

It is particularly challenging to address high frequency axion searches. In this work we have designed and tested a dielectric resonator at about 11\,GHz, which allows to maximize coupling to dark matter axions owing to its larger effective volume and high quality factor. The resonator exhibits the largest figure of merit reported in the field of axion search with resonant cavities in the range above 7\,GHz \cite{dyson2024highvolume}. It can be tuned by a clamshell mechanism that has been tested with a prototype resonator at cryogenic temperature, and we demonstrated that the quality factor is rather constant, with a mean value of 175000, throughout a tuning range of 60\,MHz. The mode mixing regions with reduced sensitivity are within few \% of the overall range, in contrast to previously reported studies in which cavities with high aspect-ratio $L_c/r_c$ operating at higher-order-modes were described as being more susceptible to mode crossings due to the high density of modes \cite{Bai_2023}. 
However, when the quality factor of the chosen mode is high the dissipative coupling with intruder modes is suppressed, leading to reduced mixing. 
Note also that tuning range can be increased to more than 2\%, as predicted by the simulations for a slightly improved cavity design. 

 The results described in this work prove that this is a valid solution to the unfavourable scaling of the scan rate with frequency, and an improved design will be used by the QUAX-$a\gamma$ collaboration to probe high axion masses.

\begin{acknowledgments}
This material is based upon work supported by INFN (QUAX experiment) and by  the U.S. Department of Energy,
Office of Science, National Quantum Information Science
Research Centers, Superconducting Quantum Materials
and Systems Center (SQMS) under the Contract No.
DE- AC02- 07CH11359.
We are grateful to E. Berto (University of Padova and INFN) who substantially contributed to the mechanical realization of this cavity and of its tuning system. A. Benato (INFN) did part of the mechanical work and M. Zago (INFN) the mechanical drawings. 
The contribution of F. Calaon and M. Tessaro (INFN) to the experiment cryogenics and electronics is gratefully acknowledged.
The cryogenic service of the Laboratori Nazionali di Legnaro provided the liquid helium to run the described experiments. 
The data that support the findings of this study are available from the corresponding authors upon reasonable request.

\end{acknowledgments}

\nocite{*}
\bibliography{mybibliography.bib}

\end{document}